\documentstyle[12pt,WSPRL]{article}

\begin{document}

\centerline{\normalsize\bf NONOBLIQUE EFFECTS AND WEAK-ISOSPIN BREAKING FROM}
\baselineskip=22pt
\centerline{\normalsize\bf EXTENDED TECHNICOLOR}
\baselineskip=16pt

\centerline{\footnotesize Guo-Hong Wu
\footnote{Talk presented at the
Top Quark Workshop, Iowa State University, Ames Iowa, May 25-26, 1995.}}
\baselineskip=13pt
\centerline{\footnotesize\it Physics Department, Yale University}
\baselineskip=12pt
\centerline{\footnotesize\it New Haven, CT 06520-8120 USA}
\centerline{\footnotesize E-mail: g\_wu@yalph2.physics.yale.edu}

\vspace*{0.9cm}
\abstracts{Several aspects of
the flavor-diagonal extended technicolor (ETC) gauge boson are reviewed.
Among them are an increase of $R_b$ that could explain the LEP $R_b$ excess,
a sizable, positive correction to the $\tau$ asymmetry parameter $A_{\tau}$,
and a contribution to the weak-isospin breaking $\rho$ parameter that is
just barely acceptable by present data.}

\normalsize\baselineskip=15pt
\setcounter{footnote}{0}
\renewcommand{\thefootnote}{\alph{footnote}}

\section{Introduction}

  In extended technicolor\cite{EL} models, the low ETC scale for
the third family often implies large corrections to the standard model
physics involving the third family of quarks and leptons.
Indeed, the third-family flavor physics could be the key to understanding the
origin of top quark mass generation and possibly of electroweak symmetry
breaking\cite{Ho}.
As the electroweak measurements have reached a precision at the one percent
level (as in $R_b \equiv \Gamma_b / \Gamma_{\mbox{\footnotesize had}}
= 0.2202 \pm 0.0020$ measured at LEP) or better
(as in $\Delta \rho_{new}= \alpha T \leq 0.4 \%$),
severe constraints can be imposed on models of dynamical electroweak
symmetry breaking\cite{Chi}.
It is therefore both timely and important to examine various models of
new physics and to see what we can learn from experiments.

 In this talk, we review the non-oblique effects
in a standard ETC\cite{CSS,CGST,Ev,Kit,Ch,W}
where the ETC bosons are standard model
singlets\footnote{Corresponding analysis for non-commuting
ETC models has also be performed\cite{CST}.}.
We will especially note that flavor-diagonal ETC exchange could explain
the LEP $R_b$ excess\footnote{A new gauge boson\cite{Ho}
existing in other models of dynamical symmetry breaking plays
a similar role.}, however the same dynamics gives a positive
contribution to the weak-interaction $\rho$ parameter which seems just
barely consistent with global fits to data.

\section{The $Zb\bar{b}$ Vertex}

   We consider a one-family TC model with technifermions belonging to
the fundamental representation of an $SU(N)_{\mbox{\scriptsize TC}}$
technicolor group and carrying the same color and electroweak quantum
numbers as their standard model counterparts.
The sideways ETC bosons mediate transitions between technifermions and
ordinary fermions, and the diagonal ETC boson couples  technifermions
(and ordinary fermions) to themselves.
The traceless  diagonal ETC generator commutes with TC and is
 normalized as $\mbox{diag}\frac{1}{\sqrt{2N(N+1)}}(1,\cdots, 1, -N)$.
The effective ETC lagrangian can thus be written as
\begin{eqnarray}
{\cal L}_{\mbox{\scriptsize ETC}} & = &
        - \frac{1}{\sqrt{2}} \sum_{i=1}^{N}
       ( X^{i,\mu}_S J_{S,i,\mu}  +  X_{S,i,\mu} J_S^{i,\mu} )
        -  X_{D,\mu} J^{\mu}_D ,
\end{eqnarray}
where $i$ is the technicolor index, and $X^{i,\mu}_S$ and $X_{D,\mu}$
stand for the sideways and diagonal ETC bosons respectively.
The sideways and diagonal ETC currents  are given by
\begin{eqnarray}
J_{S,i,\mu} & = &    g_{E,L} \bar{Q}_{iL} \gamma_{\mu} \psi_L
                   + g_{E,R}^U \bar{U}_{iR} \gamma_{\mu} t_R
                   + g_{E,R}^D \bar{D}_{iR} \gamma_{\mu} b_R ,   \\
J_S^{i,\mu} & = & (J_{S,i}^{\mu})^{\dag}  ,         \nonumber     \\
J^{\mu}_D & = &
 \frac{1}{\sqrt{2N(N+1)}} g_{E,L}
(\bar{Q}_L \gamma^{\mu} Q_L - N \bar{\psi}_L \gamma^{\mu} \psi_L)  \\
 && +  \frac{1}{\sqrt{2N(N+1)}} g_{E,R}^U
(\bar{U}_R \gamma^{\mu} U_R - N \bar{t}_R \gamma^{\mu} t_R) \nonumber \\
&& +   \frac{1}{\sqrt{2N(N+1)}} g_{E,R}^D
(\bar{D}_R \gamma^{\mu} D_R - N \bar{b}_R \gamma^{\mu} b_R) , \nonumber
\end{eqnarray}
where $Q \equiv (U,D)$ is the techniquark doublet, $\psi \equiv
(t,b)$ is the quark doublet,
and summation over color (and technicolor) indices is implied.

  To further simplify our analysis, we assume a technifermion
mass spectrum\cite{AT} where the weak scale is dominated by the nearly
degenerate techniquarks, and where the splitting between the
lighter technileptons gives a negative contribution to the $S$
parameter. We therefore have $v^2 \simeq N_C f_Q^2 \simeq (250 \mbox{GeV})^2$,
where $N_C=3$ is the number of colors, and $f_Q$ is the
Goldstone boson (GB) decay constant for the techniquark sector.
ETC corrections to the $Zb\bar{b}$ vertex are similarly dominated by
techniquarks.
For an estimate of the ETC correction to $R_b$, we only need to consider
its contribution to the left-handed $Zbb$ coupling $g_L^b$.

\subsection{Sideways ETC Exchange}

  The sideways ETC effects on $R_b$ have been discussed
previously\cite{CSS,CGST,Ev}, and we briefly review the estimate
for our one-family TC model. The relevant four-fermion operator can be
first Fierz-transformed into the product of a quark-current and a
techniquark-current,
\begin{eqnarray}
{\cal L}_{4f}^{S} & = & - \frac{g_{E,L}^2}{2m_{X_S}^2}
   (\bar{Q}_{L} \gamma^{\mu} \psi_L)
   (\bar{\psi}_{L} \gamma_{\mu} Q_L)      \nonumber  \\
 & \longrightarrow &   - \frac{g_{E,L}^2}{2m_{X_S}^2}
   \frac{1}{2N_C} \sum_{a=1}^{3}
(\bar{\psi}_{L} \gamma_{\mu} \tau_a \otimes 1_{3} \psi_L)
   (\bar{Q}_{L} \gamma^{\mu} \tau_a \otimes 1_{3} Q_L) + \cdots ,
\end{eqnarray}
where $\tau_a$'s are the Pauli matrices, $1_3$ denotes the unit matrix
in color space, and the other pieces do not contribute to the $Zb\bar{b}$
vertex.

The techniquark current can then be replaced by the corresponding
sigma model current\cite{Ge} below the TC chiral symmetry breaking scale,
\begin{equation}
\bar{Q}_{L} \gamma^{\mu} \tau_a \otimes 1_{3} Q_L  \rightarrow
i \frac{f_Q^2}{2} \mbox{Tr}(\Sigma^{\dag} \tau_a \otimes 1_{3} D^{\mu} \Sigma)
    \stackrel{\Sigma=1}{=}  - \frac{g}{c} Z^{\mu} N_C f_Q^2
        \frac{\delta^{3a}}{2}  + W^{\pm,\mu} \; \mbox{piece} ,
\end{equation}
where $\Sigma$ is the $2N_C$ by $2N_C$ exponentiated Goldstone boson matrix
transforming as $\Sigma \rightarrow L \Sigma R^{\dag}$ under
$SU(2N_C)_L \otimes SU(2N_C)_R$, $D_{\mu} \Sigma$ is its electroweak
covariant derivative, $g$ is the $SU(2)_L$ gauge coupling, and
 $c=\cos \theta_W$ ($\theta_W$ is the Weinberg angle).
The sideways ETC correction to $g_L^b$ is obtained after substituting
Eq.~(5) into Eq.~(4),
\begin{eqnarray}
\delta g_L^b(\mbox{sideways}) & = & \frac{g_{E,L}^2 f_Q^2}{8m_{X_S}^2}.
\end{eqnarray}
As this is opposite in sign to the standard model tree level value
$g_L^b = - \frac{1}{2} + \frac{1}{3} s^2$,
sideways ETC exchange decreases $\Gamma_b$ relative to the standard
model prediction.
Note that Eq.~(6) is  directly related to the TC dynamics contributing to
the weak scale, and is not dependent on the low energy effective lagrangian
approximation. The same is true for Eq.~(9).

\subsection{Diagonal ETC Exchange}

The diagonal ETC effect can be similarly analysed.
We start with the dominant four-fermion operator induced by flavor-diagonal
ETC boson exchange,
\begin{equation}
{\cal L}_{4f}^{D}  =  \frac{1}{4m_{X_D}^2} \frac{1}{N+1} g_{E,L}
      (g_{E,R}^U - g_{E,R}^D)
(\bar{Q}_R \tau_3 \gamma^{\mu} Q_R)(\bar{\psi}_L \gamma_{\mu} \psi_L),
\end{equation}
where color and technicolor summation is implied.
   Below the TC chiral symmetry breaking scale,
the right-handed techniquark current is replaced by the corresponding
sigma model current
\begin{equation}
\bar{Q}_R \tau_3 \otimes 1_3 \gamma^{\mu} Q_R  \rightarrow
i \frac{f_Q^2}{2} \mbox{Tr}(\Sigma \tau_3 \otimes 1_{3} (D^{\mu}\Sigma)^{\dag})
  \\
   \stackrel{\Sigma=1}{=}   \frac{g}{c} Z^{\mu} \frac{N_C f_Q^2}{2} .
\end{equation}
Substituting Eq.~(8) into Eq.~(7), we get the diagonal
ETC correction to $g_L^b$,
\begin{eqnarray}
\delta g_L^b(\mbox{diagonal}) & \simeq & - \frac{f_Q^2}{8m_{X_D}^2}
   \frac{N_C}{N+1} g_{E,L} (g_{E,R}^U - g_{E,R}^D) .
\end{eqnarray}

   In extended technicolor, masses of the $t$ and $b$ are given by
 $m_t \sim g_{E,L} g_{E,R}^U <\bar{U}U>$ and
 $m_b \sim g_{E,L} g_{E,R}^D <\bar{D}D>$ respectively.
 We conclude from  $m_t > m_b$ that   $g_{E,L}(g_{E,R}^U - g_{E,R}^D)>0$.
Contrary to a previous estimate\cite{Kit},
diagonal ETC exchange gives a negative
correction to $g_L^b$ and increases  $\Gamma_b$\cite{W}.

\subsection{$R_b$ Constraint}

  The total ETC correction is obtained by combining Eqs. (6) and (9),
\begin{eqnarray}
\delta g^b_{L,\mbox{\scriptsize ETC}} & \simeq & - \frac{f_Q^2}{8}
 [ \frac{g_{E,L} (g_{E,R}^U - g_{E,R}^D)}{m_{X_D}^2} \frac{N_C}{N+1}
  - \frac{g_{E,L}^2}{m_{X_S}^2} ]    \\
 & \stackrel{N=2}{\simeq} &  - \frac{v^2}{24 m_{X_S}^2}
 [ \frac{m_{X_S}^2}{m_{X_D}^2} g_{E,L} (g_{E,R}^U - g_{E,R}^D)
  - g_{E,L}^2] .        \nonumber
\end{eqnarray}
The two contributions are seen to be comparable, and
we have taken $N=2$ above as suggested by the experimental value of the
$S$ parameter.
There are of course corrections from pseudo-Goldstone-bosons (PGB's)
that need to be taken into account. These have been estimated
for QCD-like TC\cite{Ch}, and could be neglected in ETC models with
strong high momentum enhancement\cite{sETC}.

 A strong constraint can be obtained by simply requiring that the diagonal
ETC effect be as large as the effect seen at LEP.
Denoting the generic ETC couplings by $g_E$ and ETC
 boson masses by $m_{\mbox{\scriptsize ETC}}$,
we have $\delta g_{L,\mbox{\scriptsize ETC}}^b \sim - \frac{v^2}{24}
\frac{g_E^2}{m_{\mbox{\scriptsize ETC}}^2}$ from diagonal ETC.
This gives a positive correction to $R_b$,
\begin{equation}
\frac{\delta R_b}{R_b}
     \simeq  (1- R_b) \frac{2 g_L^b \delta g_L^b}
                     {{g_L^b}^2 + {g_R^b}^2}
  \sim  0.9 \% \times \frac{g_E^2}
{(m_{\mbox{\scriptsize ETC}}/{\mbox{TeV}})^2} ,
\end{equation}
where the value $s^2=0.232$ has been used. For this to agree with LEP
measurement, we need the ETC scale to be
\begin{eqnarray}
g_E^2/m_{\mbox{\scriptsize ETC}}^2 & \sim & (2 \pm 1)/\mbox{TeV}^2.
\end{eqnarray}
In strong ETC, this corresponds to
$m_{\mbox{\scriptsize ETC}} \sim 3 \ \mbox{--} \ 6 \; \mbox{TeV}$ assuming
$g_E^2/ 4 \pi^2 \simeq 1$,  and unlike QCD-like TC models\cite{CSS}
there is no simple relation between $R_b$ and $m_t$\cite{Ev}.

\section{The $\tau$ Asymmetry}

  Due to the  $1/(1-4s^2)$ enhancement, $A_{\tau}$ is particularly
sensitive to new physics. For the assumed technifermion mass spectrum,
the sideways ETC effect is negligible compared to the diagonal ETC effect,
and the ETC correction to the $Z\tau\tau$ couplings can be simply estimated,
\begin{eqnarray}
\delta g^{\tau}_{L,\mbox{\scriptsize ETC}} & \simeq & - \frac{f_Q^2}{8}
  \frac{g_{E,L}^{\tau} (g_{E,R}^U - g_{E,R}^D)}{m_{X_D}^2} \frac{N_C}{N+1}
   \\
\delta g^{\tau}_{R,\mbox{\scriptsize ETC}} & \simeq & - \frac{f_Q^2}{8}
  \frac{g_{E,R}^{\tau} (g_{E,R}^U - g_{E,R}^D)}{m_{X_D}^2} \frac{N_C}{N+1}
\end{eqnarray}
where  $g_{E,L}^{\tau}$ and $g_{E,R}^{\tau}$ are the ETC couplings for
 $\tau_L$ and $\tau_R$ respectively.

  Assuming the ETC couplings are comparable (the fermion mass
spectrum could partly arise from the hierarchy in the
technifermion condensates), and taking $N=2$ and
$g_E^2/m_{\mbox{\scriptsize ETC}}^2 \sim (2 \pm 1)/\mbox{TeV}^2$,
we have
$\delta g_{L,\mbox{\scriptsize ETC}}^{\tau} \sim
 \delta g_{R,\mbox{\scriptsize ETC}}^{\tau} \sim
 - (5.0 \pm 2.5) \times 10^{-3}$.
The ETC correction to $A_{\tau}$ is then
\begin{eqnarray}
\delta A_{\tau}/A_{\tau} (\mbox{ETC}) & \sim & 0.28 \pm 0.14
\end{eqnarray}
Note that this could be significantly reduced if $\tau$ couples
to the technifermion sector at a higher ETC scale than the $t$ quark.
Assuming $e$, $\mu$ universality, the experimental value for
$\delta A_{\tau}/A_{\tau}$ can be extracted\cite{Ho} from lepton asymmetry
measurements at LEP\cite{tau},
\begin{eqnarray}
\delta A_{\tau}/A_{\tau}(\mbox{exp}) & = & 0.14 \pm 0.10 .
\end{eqnarray}
It is seen that future lepton asymmetry measurements can have nontrivial
implications for the lepton sector in ETC.

\section{The $\rho$ Parameter}

  For the assumed technifermion mass spectrum in the one-family TC model,
there are contributions to the
weak-interaction $\rho$ parameter from the TC sector\cite{AT}, namely from
the technileptons and the PGBs.
  ETC interactions could also give a sizable correction,
and the most important ETC effect comes from the
diagonal-ETC-induced four-techniquark operator,
\begin{eqnarray}
{\cal L}_{4f}^{\Delta \rho}  & = &
   - \frac{1}{16N(N+1)} \frac{(g_{E,R}^U - g_{E,R}^D)^2}{m_{X_D}^2}
  (\bar{Q}_R \tau_3 \gamma^{\mu} Q_R) (\bar{Q}_R \tau_3 \gamma_{\mu} Q_R),
\end{eqnarray}
The leading contribution from this operator can be easily gotten
by use of Eq.~(8),
\begin{equation}
\Delta \rho_{\mbox{\scriptsize ETC}}  \simeq
   \frac{v^2}{8N(N+1)} \frac{(g_{E,R}^U - g_{E,R}^D)^2}{m_{X_D}^2}
  \simeq 0.13 \% \times  \frac{(g_{E,R}^U - g_{E,R}^D)^2}
  {(m_{X_D}/{\mbox{TeV}})^2}.
\end{equation}
And for $(g_{E,R}^U - g_{E,R}^D)^2/m_{X_D}^2 \sim
g_E^2/m_{\mbox{\scriptsize ETC}}^2 \sim (2 \pm 1)/\mbox{TeV}^2$,
this gives a correction
\begin{eqnarray}
\Delta \rho_{\mbox{\scriptsize ETC}} & \sim & (0.26 \pm 0.13) \%
\end{eqnarray}
which is barely consistent with recent global fits to
data\cite{rhoexp}.
The ETC effect on the $S$ parameter is however, negligible
compared to the TC contributions.
We refer the reader to ref.~[3] for a more complete review
on weak-isospin breaking in dynamical electroweak symmetry breaking.

\section{Conclusion}

  An ETC scale as low as
$g_E^2/m_{\mbox{\scriptsize ETC}}^2 \sim (2 \pm 1)/\mbox{TeV}^2$
is required for diagonal ETC to result in  a correction to $R_b$ as large as
seen at LEP.
This makes the diagonal ETC contribution to the $\rho$ parameter
barely acceptable.
Diagonal ETC  could also give a large and positive correction to $A_{\tau}$
if the $\tau$ couples at the same low ETC scale as the top quark.

\vspace{0.4cm}
\centerline{\bf Acknowledgments}
\vspace{0.2cm}

  I would like to thank T. Appelquist, R.S. Chivukula, N.~Evans,
W.~Marciano, S.~Selipsky and J. Terning for helpful discussions.
This work was supported by DOE under grant DE-AC02ERU3075.

\end{document}